# A hard X- ray probe to study doping-dependent electron redistribution and strong covalency in $La_{1-x}Sr_{1+x}MnO_4$

J. Herrero-Martín*, A. Mirone, J. Fernández- Rodríguez and P. Glatzel*,

*European Synchrotron Radiation Facility, F-38043 Grenoble Cedex 9 (France)*

J. Garcia and J. Blasco

*Instituto de Ciencia de Materiales de Aragón (CSIC-Universidad de Zaragoza), E-50009 Zaragoza (Spain)*

J. Geck

*IFW Dresden, P.O. Box 270116, D-01171 Dresden, Germany*

The effect of doping on the electronic structure at the Mn sites in the $La_{1-x}Sr_{1+x}MnO_4$ series (x=0, 0.3 and 0.5) was studied by means of non-resonant hard X- ray emission spectroscopy (XES). We observe a linear dichroism in the Mn Kβ main lines (3p to 1s transitions) that is strongest for x=0 and decreases with increasing x to 0.5. The Mn Kβ main lines in the poly-crystalline samples change considerably less upon increasing the hole doping (substitution of La by Sr) than it would be expected based on the change of formal valence. From this we conclude that the charge and spin density at the Mn sites are only little affected by doping. This implies that holes injected in the $La_{1-x}Sr_{1+x}MnO_4$ series mainly result in a decrease of charge density on the oxygen atoms, i.e. oxygen takes part in the charge balancing. These findings are supported by many-body cluster calculations.

## 1. INTRODUCTION

The correlated electron systems realized in transition metal oxide (TMO) materials have attracted an enormous deal of attention for their unusual and at times most puzzling electronic properties. Most prominent examples are the high-temperature superconducting cuprates or the colossal magnetoresistive manganities [1-3]. As it turns out, it is very hard to capture the unconventional electronic properties of TMO materials using conventional theoretical concepts. Many of these materials have a partially filled d-shell for example but, at the same time, exhibit insulating ground states: a clear fingerprint of strong electronic correlations. The essential step towards a better understanding of these materials is the development of effective models, which retain only the most relevant electronic degrees of freedom while still providing a realistic description of the system under study.

For instance, in doped cuprates it is well established that such a model has to contain the Cu 3d as well as the O 2p bands. It is clearly demonstrated that holes doped to the $CuO_2^-$ planes of these materials are largely localized on the oxygen ligands, forming the famous Zhang-Rice singlet with the central Cu 3d spin, which are an important ingredient to describe the charge dynamics in these materials [4].

In case of the doped manganese oxides the situation is less clear and the validity of the ionic model is still a matter of discussion [5]. However, recent X- ray absorption spectroscopy (XAS) and soft resonant X- ray scattering (RXS) studies have remarked the prominent role played by O 2p states in these materials as well [6-8].

In this paper we address this question in hole doped $La_{1-x}Sr_{1+x}MnO_4$. This series of layered compounds are iso-structural to the high temperature superconducting $La_{1-x}Sr_{1+x}CuO_4$ (space group *I4/mmm*). They do not show any magnetoresistive behavior [9] but exhibit a wide range of electrical and magnetic phases as a function of hole doping and temperature [10,11]. In particular, the existence of an integer charge segregation ($Mn^{3+}$: $Mn^{4+}$) in the ground state charge ordered phase for x=0.5 (T< 230 K) has been challenged by means of XAS and suggested that superlattice reflections have their origin on two differentiated Mn sites in the cell with a very similar associated electronic density [12]. Several theoretical calculations point in the same direction [13,14], thereby minimizing the variations in 3d orbital occupation in different Mn sites.

Nevertheless, the interpretation of XAS experiments at the transition metal K absorption main edge (1s→ 4p) in terms of the electronic structure is delicate because the spectral shape is also strongly influenced by the local coordination and geometry. The L-edge in the soft X- ray range directly probes the metal atom 3d orbitals. However, this technique suffers from other drawbacks such as surface sensitivity and the requirement of high-vacuum conditions. Furthermore, the electron configuration of the final states at the L-edge has a hole in the 2p shell with strong spin-orbit interaction and a complex multiplet structure that requires elaborate theoretical models for interpretation.

In this context, K shell X- ray emission spectroscopy (XES) is a powerful complementary technique. The hard X- ray photon- in/ photon- out is truly bulk sensitive and provides information on the density and spatial distribution of valence shell electrons in transition metal oxides [15]. More in detail, X- ray emission is a secondary process that occurs after creation of a core hole by, *e.g.*, absorption of a photon. In contrast to X- ray absorption, the emission spectra reflect the density of occupied electronic states. The Kβ main lines arise from intra-atomic 3p to 1s transitions. The hole in the 3p shell in the final state of the transition strongly interacts with the valence electrons. Slater integrals are often used to describe these interactions [16]. Amongst them, the (3p,3d) exchange dominates and splits the Kβ main lines into a sharp $K\beta_{1,3}$ and a broad Kβ' feature [17,18]. The intra-atomic origin of the strongest interactions in the Kβ main lines is well established by comparison between spectra obtained on free metal atoms and solid state systems [19].

The chemical sensitivity of the Kβ main lines arises from a modification of the (3p,3d) interaction as a result of changes in the valence shell of the Mn ion [20]. Since the exchange integral is preponderant, a high-spin/low-spin transition has a very pronounced effect [21]. The spectral changes are weaker if a high-spin configuration is maintained while the 3d orbital configuration is modified. The Kβ main lines then reflect the metal ion oxidation state. Orbital hybridization is often expressed within a cluster model by writing the metal valence shell as $\cos\alpha\,|3d^n\rangle - \sin\alpha\,|3d^{n+1}\underline{L}\rangle$ where $\underline{L}$ denotes a ligand hole and $\alpha$ describes the relative weight of the configurations [22]. A possible change of the electronic structure will modify $\alpha$ and hence the weight of the Slater integrals that shape the Kβ main lines. Structural changes will only affect the Kβ line if they are accompanied by a change of the electronic structure that changes the (3p,3d) interactions. A change of crystal field splitting, for example, may barely modify the Kβ spectra shape as shown by Peng *et al* [20].

The study of the changes in the Mn Kβ emission lines has been already carried out on other manganites of the $AMnO_3$ type (A: alkali or lanthanide). These systems belong to n=∞ in the Ruddlesden-Popper series ($La_{n(1-x)}Sr_{nx+1}Mn_nO_{3n+1}$) like $La_{1-x}Ca_xMnO_3$ [23] and $LaMn_{1-x}Co_xO_3$ [24]. However, the application to a layered, highly distorted system along the c- axis such as $La_{1-x}Sr_{1+x}MnO_4$ (n=1) offers the possibility to better evaluate the weighted contribution of the corresponding geometrical and electronic degrees of freedom.

In this paper we report the Kβ main lines spectra on poly- and single crystalline samples of the $La_{1-x}Sr_{1+x}MnO_4$ series for x=0, 0.3 and 0.5. The results are compared to recently published XANES data [12]. The different sensitivity of the two techniques allows us to disentangle the evolution of the electronic structure at the Mn sites from the local geometrical anisotropy as a function of doping. A theoretical analysis based on many-body cluster calculations is also presented [14], which sheds light on the microscopic mechanisms of observed doping dependent behavior.

## II. EXPERIMENTAL

Preparation of polycrystalline samples was done by solid state reaction from the stoichiometric amounts of $La_2O_3$, $SrCO_3$ and $MnCO_3$. The resulting powders were pressed into rods and sintered at 1500º C for 24 hours in an oxygen atmosphere. Single crystals were grown from the rods by using a homemade floating zone furnace as described in ref. [12]. They were cut with a surface normal to the [001] direction. Some pieces were ground and characterized by X- ray powder diffraction. All compounds exhibited patterns typical of a single phase. Magnetic characterization was also performed on these samples using a commercial Quantum Design SQUID magnetometer. The temperature scans of the dc magnetization for all samples were in agreement with the data reported in the literature.

X- ray emission spectra were taken at beamline ID26 of the European Synchrotron Radiation Facility (ESRF) in Grenoble (France). The incident X- rays were selected by means of a pair of cryogenically cooled Si (311) crystals. The total flux on the samples was $5 * 10^{12}$ photons/second. Non-resonant XES was performed at 6600 eV incident energy. The emission spectrometer employed one spherically bent (R = 1m) Si crystal in (440) reflection that was arranged with sample and detector (avalanche photo diode) in a horizontal Rowland geometry at 90 degrees scattering angle. All experiments were performed at room temperature. The beam size was 0.3 x 1 $mm^2$ (hor. x vert.). A slit with 1.5 mm horizontal opening was placed in front of the detector. The energy bandwidth of the spectrometer was 0.6 (0.9 eV) eV at 6490 eV for the 15 (75) degrees geometry (*vide supra*). This change in spectral broadening does not influence the results presented here. This was tested by artificially broadening the spectra and forming difference spectra. Three independent sets of data were recorded in order to determine the systematical experimental error that turned out to be ~ 0.1 eV. $LaMnO_3$ and $CaMnO_3$ were used as references. The count rates on the maximum of $K\beta_{1,3}$ line were about (2-3) *$10^4$ counts/second.

The integrated (from 6465 to 6545 eV) spectral area in all recorded emission spectra was normalized to 1. For the calculation of the integrals of the absolute values of the difference (IAD), the area below the emission curve of $LaMnO_3$ was taken as a reference. This procedure was proposed by Vankó *et al* [21] to quantify the spectral changes in $K\beta$ spectroscopy.

The geometry for the polarization dependent study of the single crystalline samples is shown in Figure 1. We consider non-resonant X- ray emission where absorption and emission are not coherently coupled. Since the core hole has s symmetry, the polarization of the incident X-ray beam is irrelevant. The emitted X- rays with moment **k'** form an angle (**k'**^c) of 15 (75) degrees with the c- axis of the crystals unit cell. This approximation to the 0 (90) degrees does not alter the results presented here. All calculated spectra are shown for the configurations purely parallel and perpendicular to the c- axis. We will refer to the two experimental geometries by **k'~//n** (**k'~⊥n**) throughout the paper because the crystal c- axis coincides with the normal **n** on the crystal surface.

We treat the emitted X- rays within the dipole approximation. Hence, in case of 3p to 1s transitions in single crystals, only 3p orbitals components that are normal to k' are observed [15,25]. It follows that the **k'~//n** geometry probes the crystal ab-plane components while **k'~⊥n** contains a 50% contribution from the crystals c- axis.

## III. THEORY

We have simulated the effect of doping along the series using a many-body cluster model from a previous work [14] which was devoted to the soft RXS at the Mn $L_{2,3}$ edge of $La_{0.5}Sr_{1.5}MnO_4$. In the model used for such work, the 3d electrons of a central Mn site are coupled to the neighbouring oxygen orbitals by a hopping term modulated by Slater-Koster parameters. Beyond the $MnO_6$ octahedra, in-plane oxygen orbitals are in turn coupled to the orbitals of the neighboring Mn sites. The Hamiltonian is written using the second quantization formalism. The Hilbert space is created by applying the Hamiltonian operator several times to a seed state that has four electrons in the 3d shell (of the central, photoexcited Mn atom), filled oxygen orbitals, and empty $e_g$ external orbitals. To keep the Hilbert space dimension below an affordable limit, only one external $e_g$ orbital (two, considering spin) directed towards the central Mn is considered. The Hilbert space built in this way allows taking the full atomic multiplet structure into account. Different configurations of the central Mn site corresponding to valence values of $Mn^{5+}$ to $Mn^{1+}$ were considered. Using a larger number of configurations does not change the computed results. We have used our model to simulate the Mn 3d occupancy as a function of doping and to simulate the Mn $K_\beta$ emission as well as its polarisation dependence.

For the Mn 3d occupancy study we have kept the same set of parameters used for the soft RXS fit in ref. [14], except for $\varepsilon_d$ in eq. 5. This term determines the energy level of the external, i.e. not photoexcited, Mn 3d orbitals and thus models in-plane oxygen mediated Mn-Mn interactions. By lowering (increasing) this level we can increase (diminish) the electron occupancy on external orbitals. The external orbitals can hence be considered as a charge reservoir that allows to mimick the effect of doping on the $MnO_6$ octahedra. Indeed, when $\varepsilon_d$ increases (decreases), the ground state energy is minimized by decreasing (increasing) the expectation value of electron occupancy on the external orbitals. As the total number of electrons in the cluster remains unchanged, such doping affects the electron occupancies on the central Mn atom and on the six neighboring oxygen atoms.

To simulate the Mn Kβ emission and its polarization dependence we have reset $\varepsilon_d$ to its RXS fit value, and added 1s and 3p shells to the model. Atomic values for the spin-orbit coupling $\zeta_{3p}$ and the Slater integrals between 3p and 3d were calculated using Cowan's Hartree-Fock code [16]. These Slater integral values were then scaled down by a factor of 0.7. The renormalised Slater integrals $F^0_{3p3d}$ and $F^0_{1s3d}$ are both taken equal to 1.1 $F^0_{dd}$. The effect of the variation of the Mn-O apical distance as a function of doping was taken into account by rescaling the Mn-O hopping [14], for the apical bond. For the emission process only the lowest-energy configurations of the system with one 1s core-hole were taken as initial states. Due to a weak 1s-3d coupling, these low-energy configurations correspond to differently coupled 1s-shell and 3d-shell spins. Higher energy configurations were not taken into account [17, 26]. In order to separate local from non-local effects, we also did calculations for an isolated $MnO_6$ octahedron by removing the external orbitals.

In the calculation of the spectra the final states in the region of the strong $K\beta_{1,3}$ line (which corresponds to S=5/2 in an ionic picture) were broadened with a Lorentzian function with width of 1.1 eV. For those of the Kβ' shoulder (S=3/2) we used 4.4 eV. These values were adjusted in order to fit the data. The reason of the larger width observed for the S=3/2 states resides in the fact that the 3d→3p radiative decay channel is favored by the spin alignment of the 3p hole with the 3d spin [17-19].

## IV. RESULTS AND ANALYSIS

### A. Single crystals

The experimental Kβ main emission lines from single crystals of $La_{1-x}Sr_{1+x}MnO_4$ (x=0, x=0.3, x=0.5) with **k'~//n** and **k'~⊥n** are shown in Figure 2. The Kβ spectrum is composed of the strong $K\beta_{1,3}$ line at high energies and a broad Kβ' feature at lower energies. The linear dichroism in a single crystal is defined as the difference between the spectra for different polarizations, i.e. in the present case different angles between the crystal axes and the direction of the emitted X- rays. The Mn-O distance along the c- axis becomes considerably shorter (~ 0.3 Å) upon increasing x from 0 to 0.5 according to neutron powder and single crystal diffraction as well as XAS studies while the average in-plane Mn-O bondlength elongates by about one order of magnitude less [12, 27]. The local geometry is tetragonally distorted for x=0 and develops towards an octahedral coordination for x=0.5 [7]. We find that the linear dichroism in the Mn X- ray emission is maximum for x=0 and minimum for x=0.5. This corresponds to the evolution of the anisotropy of their local geometry. A linear dichroism has also been observed in the X- ray absorption K main edge [12] and the question arises whether the linear dichroism has its origin in the electronic structure or the local coordination.

We show in Figure 3 the difference spectrum between polycrystalline samples of the model systems $LaMn^{III}O_3$ and $CaMn^{IV}O_3$. The shift of the $K\beta_{1,3}$ line with increasing oxidation state is accompanied by a decrease in the Kβ' intensity giving rise to a typical difference signal. Difference spectra between high-spin and low-spin Fe and Co systems have been reported by Vankó *et al* [21] and Glatzel *et al* [28] where a similar difference signal is observed. The strong similarity between the $CaMn^{IV}O_3$-$LaMn^{III}O_3$ difference curves and the linear dichroism for x=0 suggests that the latter arises from a change in electronic structure, namely a change in the valence shell orbital population.

We thus find based on the linear dichroism that for the system with the strongest tetragonal distortion (x=0) more electron density is present along the c direction than in the ab-plane. This is readily understood by the fact that the $3z^2$-$r^2$ orbital (**z** coincides with c) is the first axial 3d orbital that is populated in tetragonal symmetry with elongation along c while the $x^2$-$y^2$ orbital remains empty.

The changes of the Kβ main line along the doping series for the different polarizations (Fig. 2(b)) indicate an increase of electron density in the ab-plane and a decrease along c with Sr doping. The angle resolved emission spectra thus indicate that hole doping facilitates the charge transfer from the $3z^2$-$r^2$ to the $x^2$-$y^2$ orbitals as already proposed in ref. [7] based on XAS and XRD. It now needs to be determined by how much the integrated electron density in the Mn valence shell changes upon doping. This requires measurements on polycrystalline samples as discussed in the next section.

**B. Polycrystalline samples**

The Kβ main emission spectra on polycrystalline samples of $LaSrMnO_4$, $La_{0.7}Sr_{1.3}MnO_4$ and $La_{0.5}Sr_{1.5}MnO_4$ are shown in Figure 4 together with the reference systems $LaMnO_3$ and $CaMnO_3$. It has been shown by several authors that the strong $K\beta_{1,3}$ peak moves to lower energies and the Kβ' feature decreases in intensity with lower spin state [15,20]. The spin state translates to oxidation state in half-filled shells if a high- spin configuration is maintained. The expected spectral change is observed between $Mn^{III}$ in $LaMnO_3$ and $Mn^{IV}$ in $CaMnO_3$. We followed the procedure as suggested by Vankó *et al* [21] to quantify the spectral changes by means of the integrals of the absolute values of the difference spectra (IAD). The results are shown in Figure 5.

Within a simple ionic picture one would assign the spectral change from $LaMnO_3$ to $CaMnO_3$ to an integer decrease of charge on Mn. Formally, the spectral changes in the $La_{1-x}Sr_{1+x}MnO_4$ series should span half of the change in the reference systems. This is based on the assumption that only Mn takes part in balancing of the charge upon hole doping. The charge of the oxygen ions is always assumed to be -2 in the ionic picture.

However, the Mn Kβ main line spectral shapes as well as the IAD values are nearly identical for the three samples of the $La_{1-x}Sr_{1+x}MnO_4$ series. The change in IAD value between x=0 and x=0.5 is approximately 15% of the change between $LaMnO_3$ and $CaMnO_3$ as opposed to the expected 50%. The electronic structure at the Mn sites therefore changes much less than it would be anticipated based on an ionic model.

We also show in Fig. 5 the energy position of the Mn K main absorption edge (taken as the first maximum of the derivative of the absorption spectra) from the XANES measurements reported in ref. [12] on polycrystalline samples of this series of samples. The main edge shifts within the $La_{1-x}Sr_{1+x}MnO_4$ series considerably more than what is observed in the Kβ emission lines. While the K-edge position follows approximately what would be expected based on the formal valence, the spectral changes in the Kβ lines are considerably less than expected.

This apparent contradiction can be reconciled by noting that the K absorption edge is sensitive to both, the electronic structure and the local coordination while the Kβ lines are mainly sensitive to the electronic structure. We thus conclude that the edge shift in the in the $La_{1-x}Sr_{1+x}MnO_4$ series mainly arises from changes of the inter-atomic distances and that the electronic structure on the Mn ion changes only very little as evidenced by Kβ spectroscopy. As a consequence, we find that the injected holes are mainly localized on the oxygen lattice as it was suggested by Ferrari *et al* [29]. We also point to the fact that the Kβ main line for $LaSrMnO_4$ is already markedly different from $LaMnO_3$ even though both compounds formally contain $Mn^{III}$ ions. Layered and cubic structures thus give a different electron configuration at the Mn site. In fact, Mn in $LaSrMnO_4$ appears more oxidized than in $LaMnO_3$.

The experimental results are supported by cluster calculations. In general, a satisfactory agreement between experiment and theory can be achieved for the non-resonant X- ray emission lines [15,20,22]. The reason is that the Kβ lines are not sensitive to the fine structure of the energy levels in the valence shell. Correct treatment of the electron-electron interactions within an atomic multiplet model already provides good agreement. Including the ligand environment mainly serves for a correct determination of the valence shell electron occupation.

In Figure 6(a) we show the occupancy variations for the central photoionized Mn 3d (solid line), the O 2p (dotted line) and external orbitals (dashed line), as a function of the difference in energy of 3d orbitals in the external ($\varepsilon_{d'}$) and internal ($\varepsilon_d$) Mn atoms as established in [14]. When the external orbital energy increases, electronic charge of these orbitals moves to the central $MnO_6$ cluster. The cluster calculation shows that the central Mn 3d occupancy variation is less than 0.1 electrons while the total charge on oxygen 2p orbitals, in the $O_6$ octahedra, varies by about 0.5 electrons. The theoretical model therefore predicts that the effect of doping affects mainly the occupancy of oxygen orbitals while the variation of the charge on the Mn ions remains small.

Figure 6 (b) shows the calculated XES spectra for different values of the tetragonal distortions as described by $\delta$, where $(1+\delta)$ is the ratio between c- axis and in-plane Mn-O bond lengths. We have set equal distances for the a and b directions, and we have rescaled the hopping parameters accordingly. It can be observed that the anisotropy, in particular the separation between the main peaks for in-plane and perpendicular polarizations, has the correct order of magnitude. We performed the same calculation for an isolated atom (setting the Mn-O hybridisation to zero) and using a crystal field to remove the degeneracy in the 3d orbitals (spectra not shown). This calculation strongly overestimates the linear dichroism and the agreement with experiment in this case is very poor.

The reason is that in the non-hybridized case the 3d $3z^2 -r^2$ orbital has an occupation equal to one in presence of a crystal field while in the hybridized model the occupation is less than one for the 3d $3z^2 -r^2$ orbitals. This is due to the hopping between O and external orbitals, and between Mn and O. For the same reason the 3d $x^2-y^2$ occupation is non-zero in presence of hybridization. Such effect reduces the charge anisotropy around the Mn ion.The linear dichroism has the correct sign for $\delta=0.2$ with the peak separation decreasing when $\delta$ is reduced. This trend in the calculations is the result of the competition between two effects: first, the c-axis distortion which tends to increase the charge density along the c- axis, and second, a kinetic effect which favors in-plane occupancy to lower the energy due to the oxygen-mediated hopping to external Mn orbitals. This conclusion is verified by a calculation for an isolated $MnO_6$-octahedron where the external orbitals have been removed. In this case, the anisotropy is obtained immediately as soon as the cubic symmetry is broken and remains practically independent of $\delta$. The model therefore implies that increase of the in-plane occupancy is related to intersite-hopping processes within the ab-plane.

Our calculations give, for $\delta$ going to zero, a lower $K\beta_{1,3}$ peak energy for c-axis than for in-plane polarization. The experiment shows instead almost identical positions for the two peaks. We think that the reason of this discrepancy lies in the limited size of our model. In reality, the ground state in manganites is realised by a symmetry-broken phase which consists of different non-equivalent Mn sites. We think that in the real system some of these sites conserve a charge anisotropy with more charge along the c-axis which counterbalance the effect of those sites whose charge has been reoriented in plane by Mn-Mn interaction. We think that a more realistic calculation, done taking into account a bigger cluster, would give a better agreement. Such calculation within the present model is at the moment limited by currently available computing power.

## V. CONCLUSIONS

In summary, we have presented a non-resonant K-shell X- ray emission spectroscopic study for different compositions in the $La_{1-x}Sr_{1+x}MnO_4$ series of layered manganites, both on poly- and single crystalline samples. Kβ main emission lines from powders show that the electron density localized on Mn atoms appear to be rather constant along the series for $0 \leq x \leq 0.5$. A similar behavior was also reported in $La_{1-y}Ca_yMnO_3$ for $0<y<0.3$ [23], where it was related to the transition from the insulator to conductive character of the samples that takes place in this doping range. This result is surprising because the formal valence for these compositions evolves from $Mn^{3+}$ for x=0 to $Mn^{3.5+}$ when x=0.5. This points out to an active role of the neighboring O atoms in the charge- transfer process and an almost unchanged Mn 3d occupancy.

In addition the polarization dependent measurements done on single crystals reveal that the main change related to Mn 3d states is a redistribution of the $e_g$ charge from states oriented out-of-plane to states parallel to the ab-plane.

These two main results, namely (i) the small change of the Mn valence upon increasing x and (ii) the redistribution of the $e_g$ electrons have been reproduced by many-body cluster calculations. They yield a pronounced O 2p character of the doped charge carriers, which explains (i). Further, the extended cluster calculation implies that the charge redistribution related to the $e_g$ orbitals is mainly caused by non-local effects (kinetic energy, exchange interactions) within the ab-planes.

These conclusions are in good agreement with a previous O K edge study [7], where complementary results regarding the unoccupied O 2p states were reported.

Taken together the above discussion shows that the doping induced changes to the electronic structure involve both the oxygen 2p and the Mn 3d bands, which both play an active role, and contrast with the classical ionic model perspective as previously stated [6, 8, 30, 31]. Finally, this result is an important input for the development of effective models which is an essential step towards a better understanding of these complex materials.

## Acknowledgements

We wish to thank the ESRF for beam granting and financial support from Spanish MICINN, FIS2008-03951. J. G acknowledges the German DFG through the Emmy-Noether Program J. F.-R. the Gobierno del Principado de Asturias for the financial support from Plan de Ciencia, Tecnología e Innovación PCTI de Asturias 2006–2009.

**Figure captions**

Figure 1. Geometry of the polarization dependent XES measurements on single crystalline samples in the **k'~//n** (in-(ab) plane) and **k'~⊥n** (out-of-plane) configurations. **E'** refers to the electric vector of the non- polarized emitted radiation. The crystal c- axis coincides with **n.**

Figure 2. Kβ main emission from single crystals with dopings x=0 (blue), 0.3 (green) and 0.5 (red), corresponding to in- and out-of- plane polarization configurations. Solid and dotted lines refer to **k'~//n** and **k'~⊥n** spectra, respectively. In (a) spectra are arranged by compound and in (b) by emission angle. The inset shows the maxima of the $K\beta_{1,3}$ lines. Spectra have been vertically shifted for the sake of clarity.

Figure 3. Linear dichroic signal of $LaSrMnO_4$ (open circles) compared to the subtraction of $LaMnO_3$ to $CaMnO_3$ powder emission spectrum (closed circles).

Figure 4. Mn Kβ main emission lines from polycrystalline samples of $La_{1-x}Sr_{1+x}MnO_4$ for x=0 (blue), 0.3 (red) and 0.5 (green). Corresponding spectra from $CaMnO_3$ (dotted) and $LaMnO_3$ (dashed) are also shown for comparison. The inset shows in detail the $K\beta_{1,3}$ lines.

Figure 5. Integrals of the absolute values of the difference spectra (IAD) for the Mn Kβ main emission lines from polycrystalline samples with x=0, 0.3 and 0.5, and $CaMnO_3$ (circles). All of them are relative to $LaMnO_3$. For comparison, the shift of the energy position at the Mn K absorption spectra in the series is also shown (crosses). The scale of the ordinate is chosen such that the IAD values and edge position coincide for $LaMnO_3$ and $CaMnO_3$.

Figure 6. (a) Calculation of the occupancy variation for the 3d shell of the central photoionized Mn (solid line), 2p of nearest six oxygens (dotted line) and external orbitals (dashed line) as a function of the $\varepsilon_d$ parameter; (b) main emission lines calculated for **k'//n** (solid) and **k'⊥n** (dotted) as a function of the tetragonal distortion of $MnO_6$ octahedra.

*Figure 1*

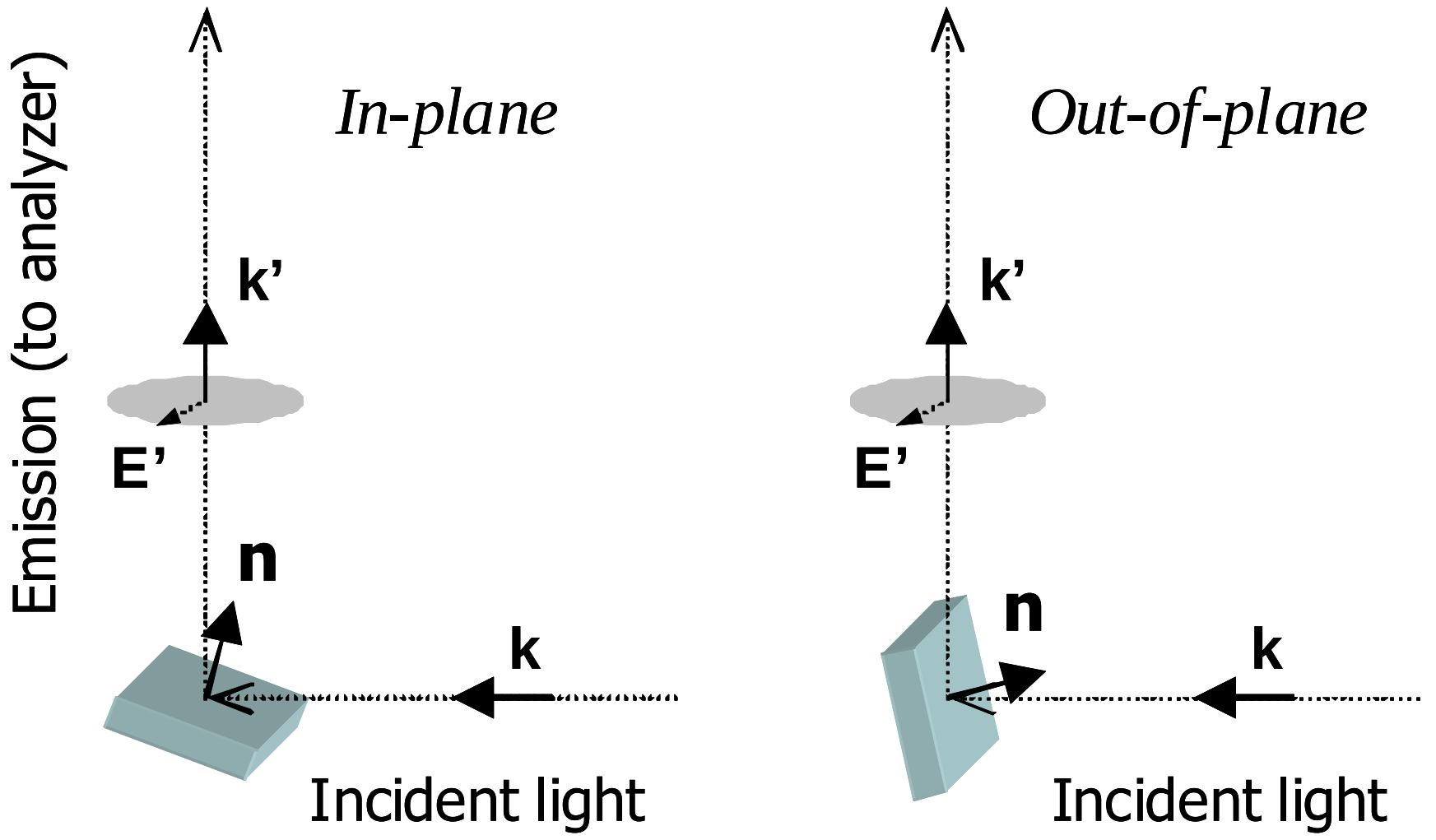

*Figure 2*

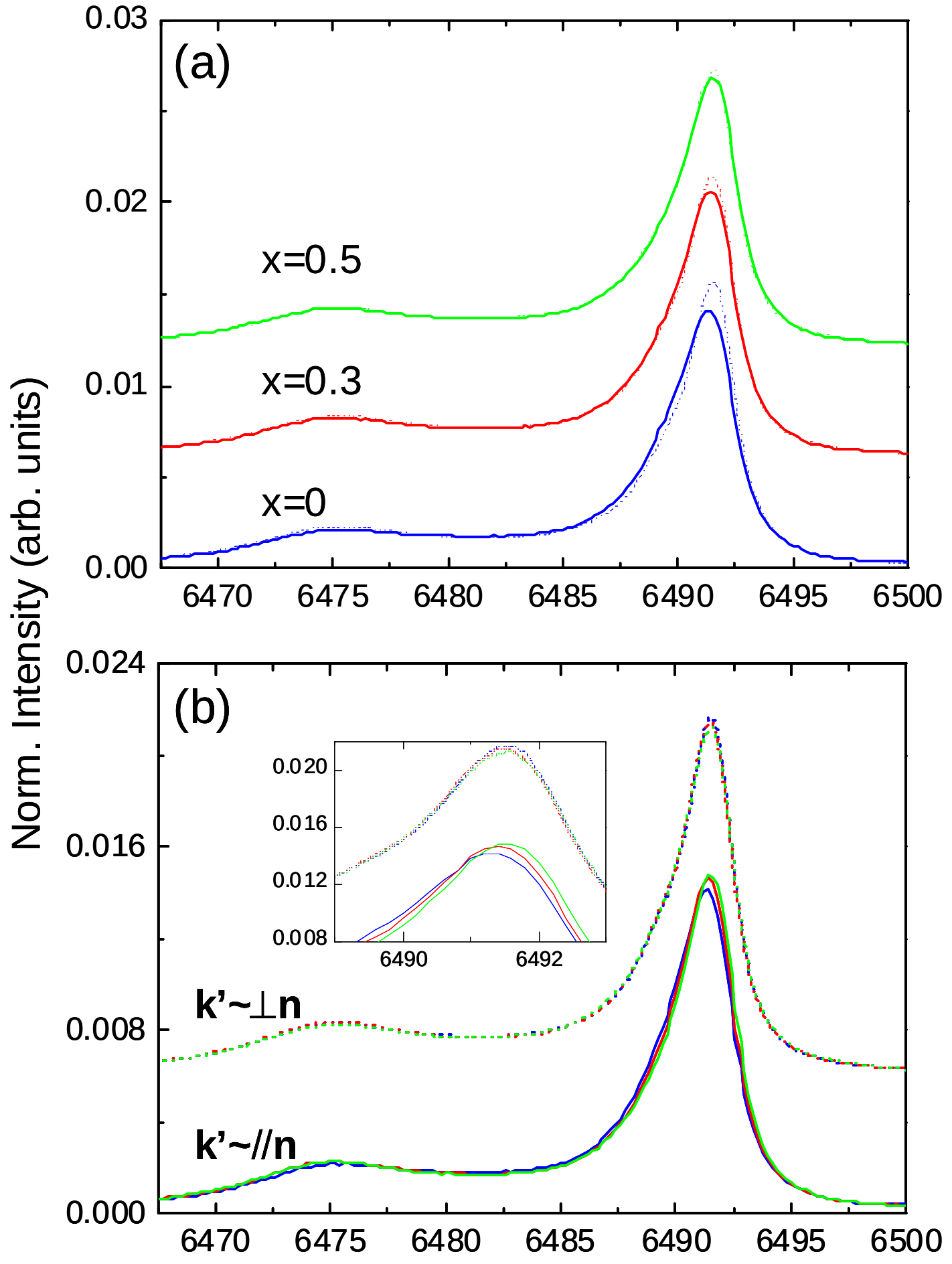

*Figure 3*

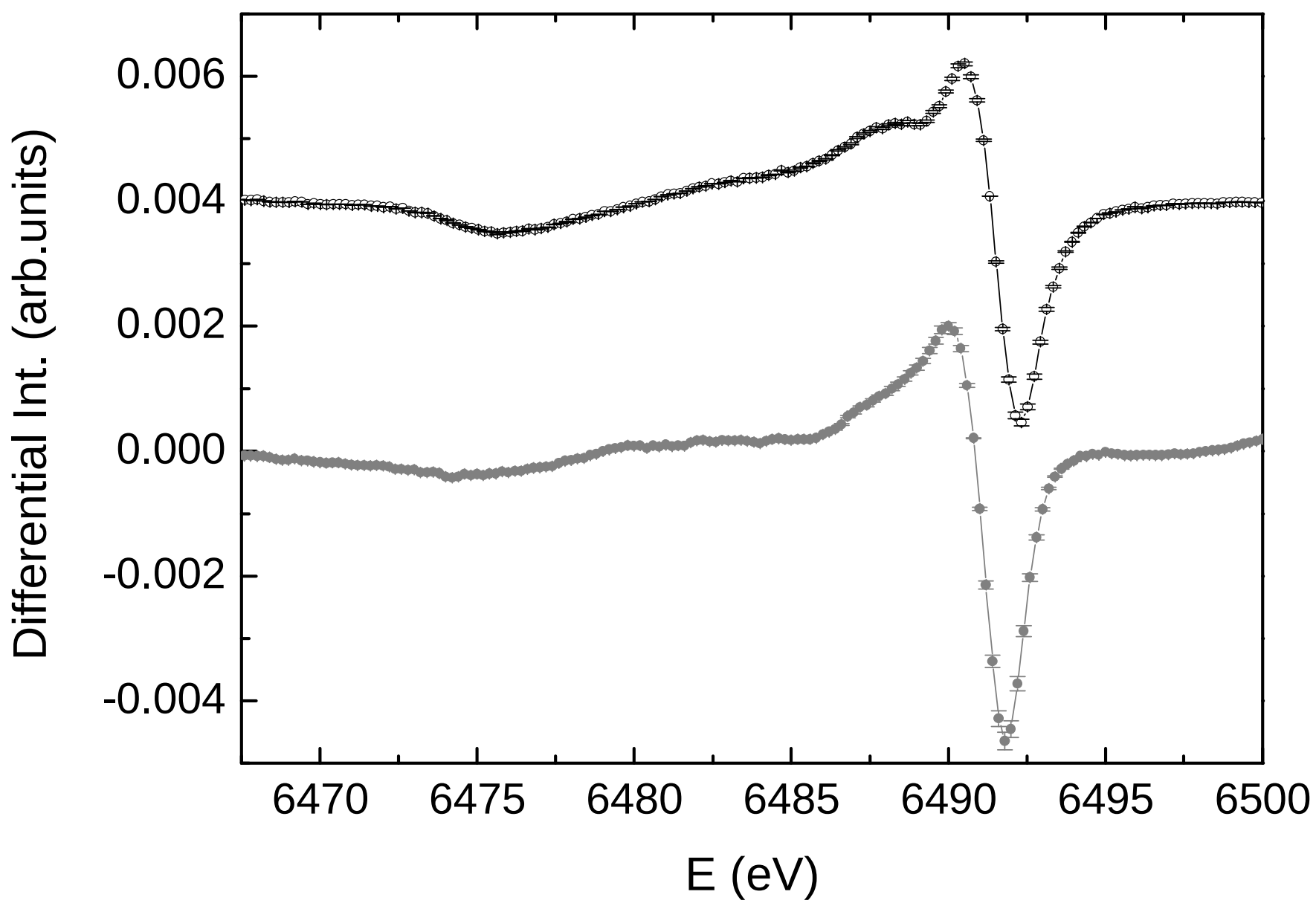

*Figure 4*

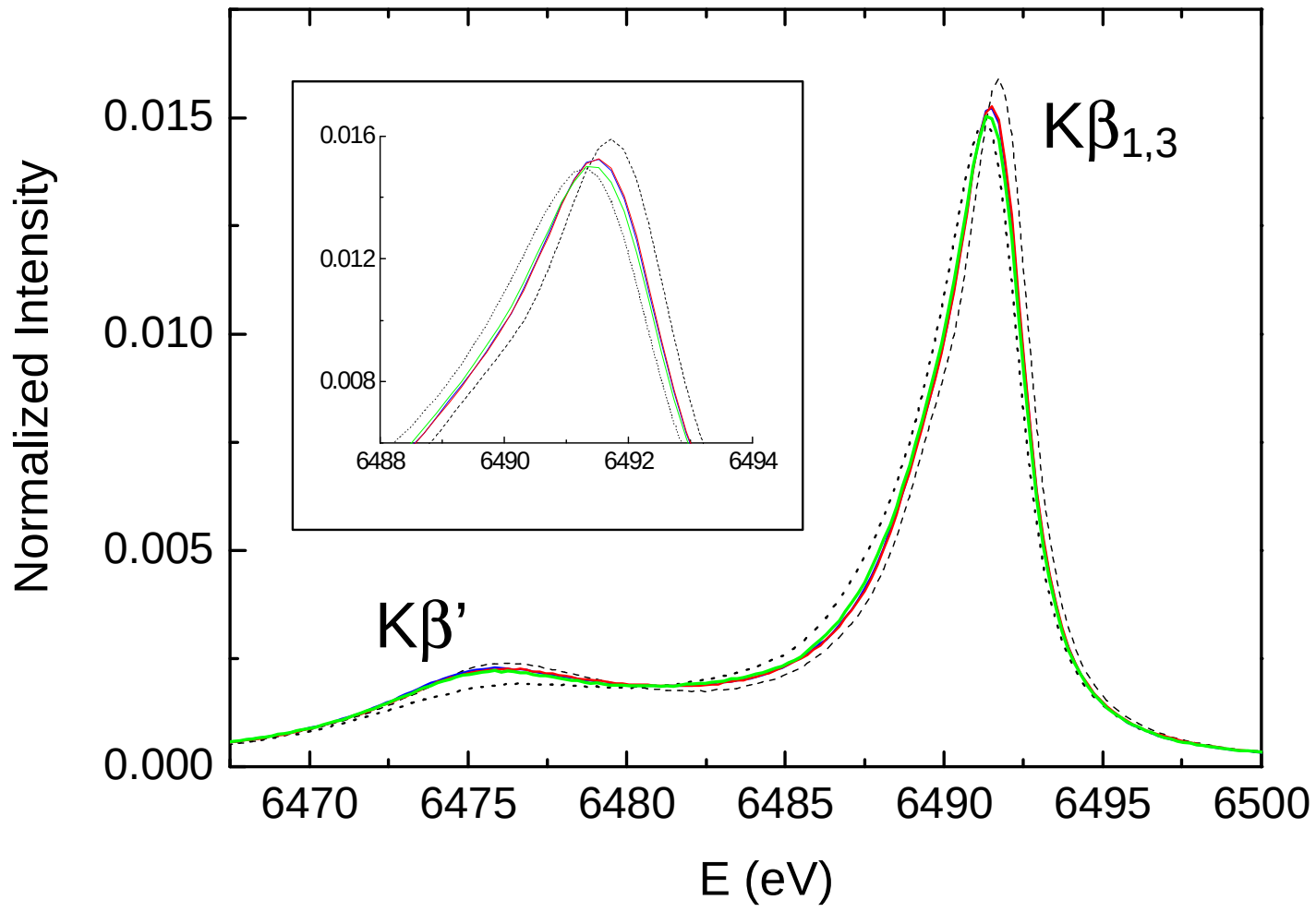

*Figure 5*

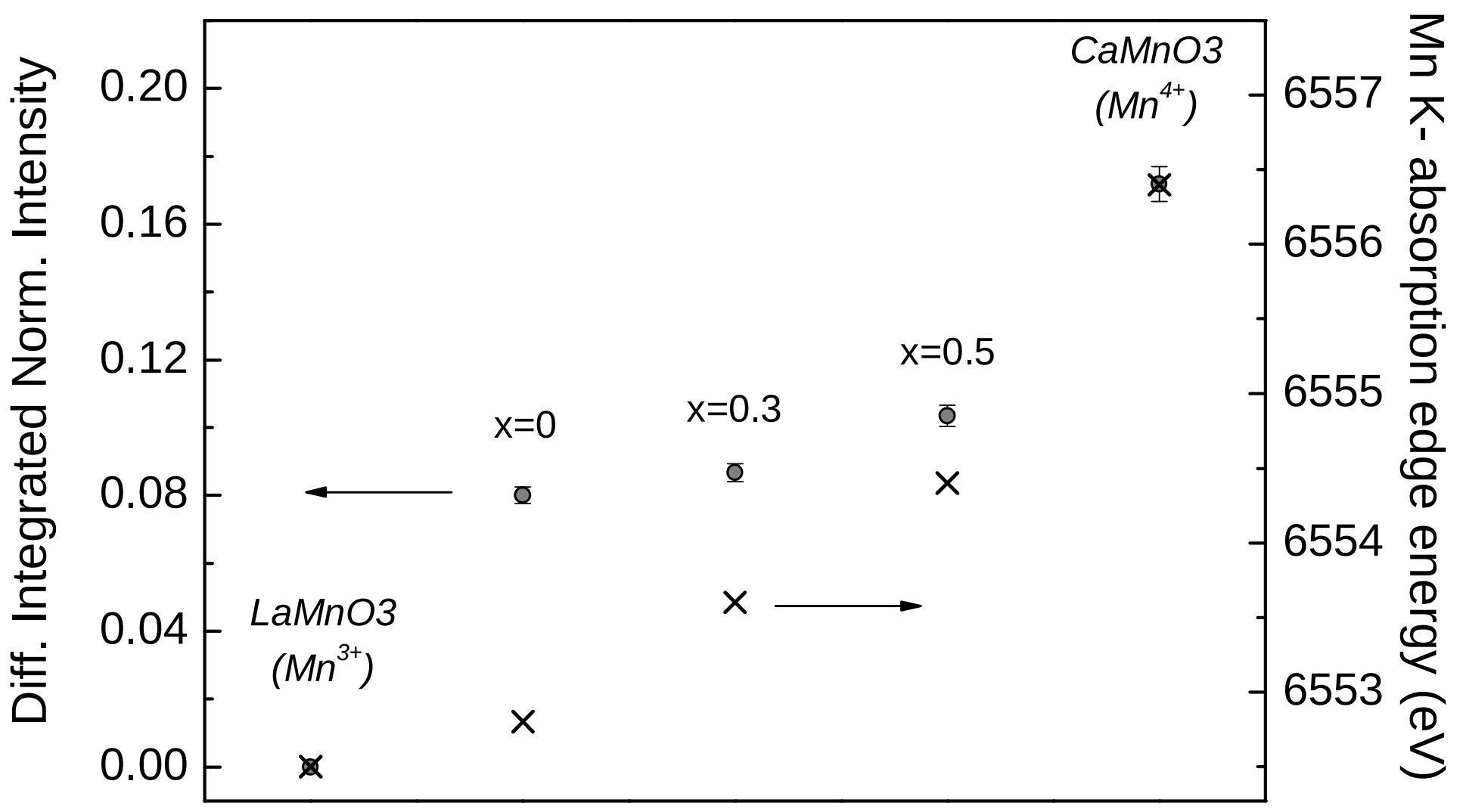

*Figure 6*

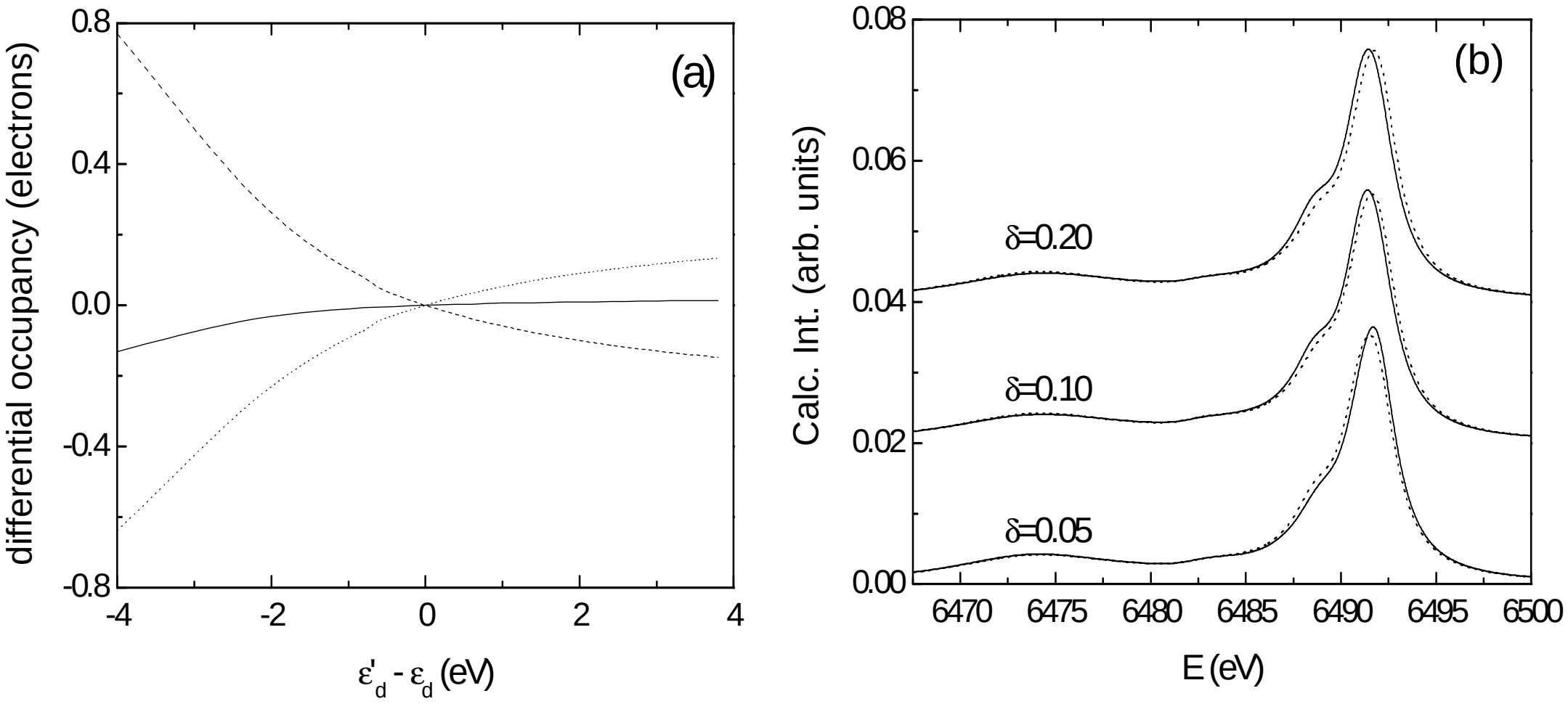